\documentclass[sigconf,screen]{acmart}
\usepackage{subcaption}
\acmSubmissionID{1}
\usepackage{todonotes}
\usepackage{url}
\usepackage{enumitem}
\usepackage{graphicx}
\usepackage{caption}
\usepackage{amsmath}
\usepackage{tabu}
\usepackage{multirow}
\usepackage{hyperref}
\usepackage{adjustbox}
\usepackage{mathptmx}
\usepackage{amsfonts}
\usepackage{algpseudocode}
\usepackage{pgfplots}
\usepackage{gensymb}
\pgfplotsset{compat=1.12}
\usepackage{soul}
\usepackage{footnote}
\usepackage{lipsum}
\usepackage[ruled,vlined, linesnumbered]{algorithm2e}
\usepackage{subcaption}
\usepackage{listings}
\colorlet{punct}{red!60!black}
\definecolor{background}{HTML}{FFFFFF}
\definecolor{delim}{RGB}{20,105,176}
\colorlet{numb}{magenta!60!black}

\newcommand*\circledw[1]{\tikz[baseline=(char.base)]{
            \node[shape=circle,draw,inner sep=0.75pt, text=black,fill=white] (char) {#1};}}
\lstdefinelanguage{json}{
    basicstyle=\footnotesize\ttfamily,
    numbers=left,
    numberstyle=\scriptsize,
    xleftmargin=2.3em,
    xrightmargin=0.5em,
    framexleftmargin=1.9em,
    stepnumber=1,
    numbersep=8pt,
    showstringspaces=false,
    breaklines=true,
    frame=single,
    backgroundcolor=\color{background},
    literate=
     *{0}{{{\color{numb}0}}}{1}
      {1}{{{\color{numb}1}}}{1}
      {2}{{{\color{numb}2}}}{1}
      {3}{{{\color{numb}3}}}{1}
      {4}{{{\color{numb}4}}}{1}
      {5}{{{\color{numb}5}}}{1}
      {6}{{{\color{numb}6}}}{1}
      {7}{{{\color{numb}7}}}{1}
      {8}{{{\color{numb}8}}}{1}
      {9}{{{\color{numb}9}}}{1}
      {:}{{{\color{punct}{:}}}}{1}
      {,}{{{\color{punct}{,}}}}{1}
      {\{}{{{\color{delim}{\{}}}}{1}
      {\}}{{{\color{delim}{\}}}}}{1}
      {[}{{{\color{delim}{[}}}}{1}
      {]}{{{\color{delim}{]}}}}{1},
}

\DeclareRobustCommand*{\IEEEauthorrefmark}[1]{\raisebox{0pt}[0pt][0pt]{\textsuperscript{\footnotesize\ensuremath{\ifcase#1\or *\or \dagger\or \ddagger\or%
    \mathsection\or \mathparagraph\or \|\or **\or \dagger\dagger%
    \or \ddagger\ddagger \else\textsuperscript{\expandafter\romannumeral#1}\fi}}}}

\AtBeginDocument{%
  \providecommand\BibTeX{{%
    \normalfont B\kern-0.5em{\scshape i\kern-0.25em b}\kern-0.8em\TeX}}}

\copyrightyear{2023} 
\acmYear{2023} 
\setcopyright{acmlicensed}\acmConference[SC-W 2023]{Workshops of The International Conference on High Performance Computing, Network, Storage, and Analysis}{November 12--17, 2023}{Denver, CO, USA}
\acmBooktitle{Workshops of The International Conference on High Performance Computing, Network, Storage, and Analysis (SC-W 2023), November 12--17, 2023, Denver, CO, USA}
\acmPrice{15.00}
\acmDOI{10.1145/3624062.3624271}
\acmISBN{979-8-4007-0785-8/23/11}

\begin{document}

\title{Sustainability in HPC: Vision and Opportunities
}
\author{Mohak Chadha$^{1}$, Eishi Arima$^{1}$, Amir Raoofy$^{1,2}$, Michael Gerndt$^{1}$, Martin Schulz$^{1,2}$
}
\affiliation{%
\institution{$^1$\{firstname.lastname, martin.w.j.schulz\}@tum.de, Technische Universit{\"a}t M{\"u}nchen 
\country{Germany}}
}
\affiliation{%
\institution{$^2$\{firstname.lastname\}@lrz.de, Leibniz-Rechenzentrum
\country{Germany}}
}
\renewcommand{\authors}{Mohak Chadha, Eishi Arima, Amir Raoofy, Michael Gerndt, Martin Schulz}
\renewcommand{\shortauthors}{Mohak Chadha, Eishi Arima, Amir Raoofy, Michael Gerndt, Martin Schulz}





\begin{abstract}
Tackling climate change by reducing and eventually eliminating carbon emissions is a significant milestone on the path toward establishing an environmentally sustainable society. As we transition into the exascale era, marked by an increasing demand and scale of HPC resources, the HPC community must embrace the challenge of reducing carbon emissions from designing and operating modern HPC systems. In this position paper, we describe challenges and highlight different opportunities that can aid HPC sites in reducing the carbon footprint of modern HPC systems.

\end{abstract}



\maketitle


\section{Introduction}
\label{sec:intro}
The combustion of fossil fuels for electricity production is a prominent driver of global greenhouse gas (GHG) emissions, which has far-reaching implications for climate change on a planetary scale~\cite{fossilfuels}. According to a recent report~\cite{ipcc_report} from the U.N. Intergovernmental Panel on Climate Change (IPCC), the world is rapidly exhausting the Paris Agreement goal~\cite{parisagreement} of limiting the Earth's average temperature increase to 1.5\degree C by the end of the 21st century. Exceeding this limit for a sustained duration can lead to irreversible environmental damage and an increased probability of climate catastrophes~\cite{climategoal}.  A significant amount of the world's total electricity is consumed by datacenters and HPC systems today~\cite{masanet2020recalibrating}. For instance, the exascale system \textit{Frontier} at the Oak Ridge National Lab consumes $20$MW of power in continuous operation, while the upcoming \textit{Aurora}~\cite{aurora} system at the Argonne National Lab is estimated to draw $60$MW~\cite{sustainablesuper}. With the growing need for HPC resources, increasing migration to the cloud, and expanding demand from novel domains such as Generative AI~\cite{chatgpt}, datacenters are projected to consume between $8$\% to $13$\% of the world's total electricity by 2030~\cite{Andrae2015OnGE}. 



Characterizing the carbon impact of an HPC system cannot be accurately done by only measuring its power or electricity consumption. Its carbon footprint spans multiple sources of emissions, which are classified into three scopes as defined in the GHG protocol~\cite{ghgprotocol}. Scope 1 emissions include \emph{on-site} emissions resulting from the direct burning of fuels or chemicals for power generation, as well as the emissions associated with the activities of the center's staff. Scope 2 emissions derive from \emph{purchasing} energy from the electric grid to power up the HPC system. On the other hand, Scope 3 emissions originate from the carbon embedded in \emph{upstream activities} in manufacturing the various components, such as CPUs, GPUs, memory, networking interconnects, and storage, before they are deployed into operation. Within HPC, the \textit{operational} carbon emissions include both Scope 1 and Scope 2 emissions, while the \textit{embodied} carbon emissions are encapsulated within Scope 3~\cite{gupta2021chasing}. Prior research~\cite{mythsembodied} and statements from commercial cloud providers~\cite{envsusmicrosoft} have indicated that Scope 1 emissions are negligible\footnote{Exceptions do exist, such as RIKEN, where on-site fossil fuel based power generation might lead to significant Scope 1 emissions.} as compared to Scope 2 and Scope 3 emissions. Towards this, in this paper, we highlight and present different opportunities for reducing the embodied (\S\ref{sec:embodiedcarbon}) and operational carbon footprint (\S\ref{sec:opcarbon}) of HPC systems.

\section{Reducing Embodied Carbon}
\label{sec:embodiedcarbon}
\begin{figure}[t]
\centering
\includegraphics[width=\columnwidth]{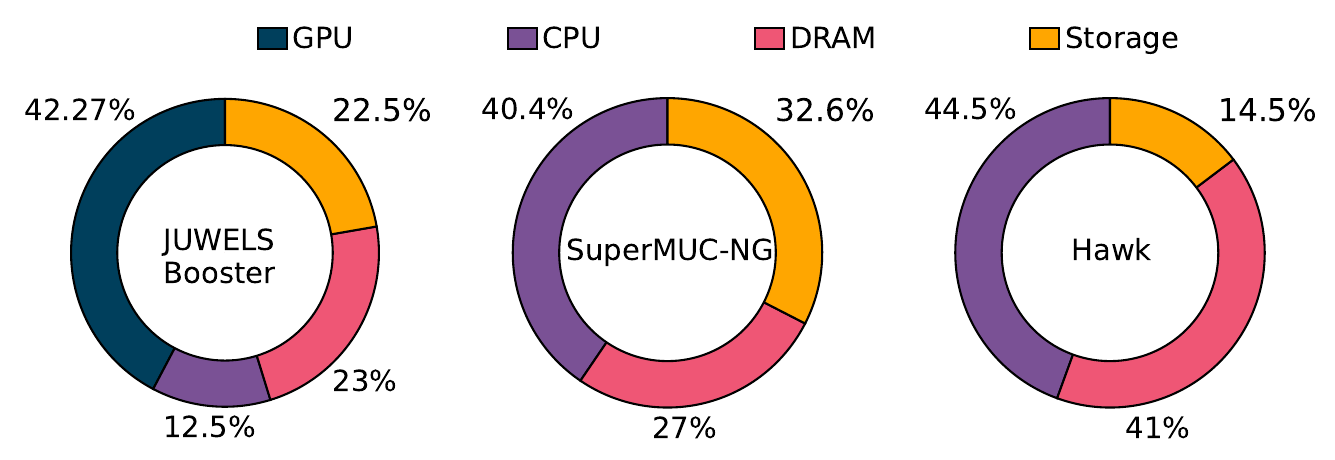}
\caption{Estimated embodied carbon footprint contribution from the different components~\cite{li2023sustainable} in the Top-3 HPC systems in Germany~\cite{top500}.}
\label{fig:embodiedcarbon}
\end{figure}

As supercomputers continue to scale in size~\cite{top500perf}, incorporating an extensive array of heterogeneous components within individual nodes~\cite{futurehpc}, it is essential for the system architects to be aware of the carbon footprint that the different hardware components \textit{embody}. More specifically, the \textit{embodied carbon} refers to the carbon emissions associated with non-recurring expenditures encompassing a spectrum of activities such as production, transportation, and disposal of the components and equipment used in the HPC system and the site. Prior research~\cite{gupta2022act} has shown that the production phase dominates the total embodied carbon emissions and constitutes \textit{manufacturing} and \textit{packaging} carbon. The former represents emissions generated from producing components such as silicon wafers, transistors, and interconnects, while the latter represents emissions from integrating the produced components into functional chips.

Fig.~\ref{fig:embodiedcarbon} shows the embodied carbon footprint contributions from the different hardware components for the top three HPC systems in Germany, i.e., \textit{Juwels Booster}~\cite{juwelsbooster}, \textit{SuperMUC-NG}~\cite{supermucng}, and \textit{Hawk}~\cite{hawk}. To model the carbon impact from the different components, we use the methodology presented in Li~et~al.~\cite{li2023sustainable}. The Juwels Booster system consists of $3744$ Nvidia A100 GPUs, $1872$ AMD EPYC 7402 processors, $0.47$PB of DRAM, and $37.6$PB of storage, while SuperMUC-NG and Hawk are CPU-only systems with $12960$ Intel Skylake and $11264$ AMD Rome processors, respectively. SuperMUC-NG has $0.72$PB of DRAM with $70.26$PB of storage, while Hawk has $1.4$PB of DRAM with $42$PB of storage. Due to the lack of production carbon-emission reports, we omit the embodied carbon footprint contributions from high-performance networking interconnects that are integral components within HPC systems. Reaffirming the results from Li~et~al.~\cite{li2023sustainable}, we observe that GPUs have a significantly higher carbon embodied footprint than the others. 
This can be attributed to the larger die area of GPUs leading to higher manufacturing carbon emissions. In addition, the collective embodied carbon footprint for memory and storage adds up to a significant fraction of the footprint of the entire HPC system. For instance, memory and storage account for $43.5$\%, $59.6$\%, and $55.5$\% embodied carbon emissions for the three systems, respectively. 


The share of embodied carbon within the overall carbon impact of an HPC system (\S\ref{sec:intro}) varies based on its geographical location and the energy source used for its operation. For instance, the LRZ Supercomputing centre~\cite{lrz} situated in Garching operates exclusively on hydropower, resulting in a relatively low carbon intensity (\S\ref{sec:opcarbon}) of $20$ gCO2/kWh, in contrast to data centers relying on non-renewable energy sources like coal which has a significantly higher carbon intensity of $1025$ gCO2/kWh~\cite{gupta2022act, coal}. Therefore, for LRZ, embodied carbon emissions dominate the overall carbon footprint. As a rule of thumb~\cite{mythsembodied}, for data centers operating with $70-75$\% renewable energy, the embodied carbon accounts for $50$\% of the total carbon emissions. To this end, opportunities for reducing the embodied carbon emissions of an HPC system can be divided into three categories: \circledw{1} hardware architecture design (\S\ref{sec:chip-level}), \circledw{2} system design (\S\ref{sec:system-level}), and \circledw{3} operational strategy (\S\ref{sec:lifetimerr}).

\subsection{Designing Carbon-efficient HPC Components}
\label{sec:chip-level}

Supercomputers have been typically composed of commodity hardware,  
however designing and customizing chips for target applications (i.e., co-design) has become more significant recently. 
As an example, the European Processor Initiative (EPI) was established and got funded in 2018 in order to develop general purpose CPUs and accelerators optimized for HPC and autonomous systems~\cite{gagliardi2019international}. 
Another example is the A64FX processor that powers the supercomputer Fugaku in Japan and was developed throughout their thorough co-design efforts~\cite{fugaku}. 
The trend will also be driven by the rapidly growing open RISC-V community~\cite{risc-v}, and we foresee that more own custom  processors will appear in future HPC systems. 

When developing a custom HPC processor, the carbon efficiency 
will be one of the first-class citizen in addition to performance and energy efficiency metrics. 
A recent study presented an architectural framework to assess the carbon footprint for microprocessors and demonstrated that the optimal design point could change depending on the design objective metric such as CDP (Carbon Delay Product), CEP (Carbon Energy Product), and others~\cite{gupta2022act}. 
These carbon-efficiency metrics depend on the semiconductor fabrication, the technology generation, as well as the carbon intensity of the power grid at which the processor will operate (as both the embodied and operational carbons need to be considered). 
Moreover, recent HPC processors are typically composed of multiple chiplets, which are integrated via the 2.5D silicon interposer technology, and they can include different modules manufactured by different fabrications. 
For instance, Intel's Ponte Vecchio GPU consists of 63 chiplets, manufactured with five different technology nodes~\cite{gomes2022ponte}. 
\textit{Under these conditions, carbon-aware HPC processors will have to be designed and optimized in an 
end-to-end manner: (1) assessment for the typical carbon intensity of the power grid where the processor will operate; (2) package-level optimization to decide the combination of chiplets as well as their fabrications; and (3) design space exploration for each chiplet under the given fabrication and assuming carbon intensity of the power grid. }


\begin{table}[t]
\caption{Recent modern HPC systems at LRZ~\cite{lrz}.}
\begin{adjustbox}{width=0.7\columnwidth,  center}
\begin{tabular}{|c|c|c|}
\hline
HPC System          & Start of Operation & Decomissioned \\ \hline
SuperMUC~\cite{lrzdecommisioned}            & 2012               & 2018          \\ \hline
SuperMUC Phase 2~\cite{lrzdecommisioned}    & 2015               & 2019          \\ \hline
SuperMUC-NG~\cite{supermucng}         & 2019               &   2024           \\ \hline
SuperMUC-NG Phase 2~\cite{supermucngphase2} & 2023               &   -           \\ \hline
ExaMUC~\cite{examuc}             & 2025               &   -            \\ \hline
\end{tabular}
\end{adjustbox}

\label{tab:hpcsystems}
\end{table}

\subsection{System Architecture and Procurement}\label{sec:system-level}
The increasing demands for reducing the carbon emissions of supercomputers will affect also the design decisions at the system architecture level. 
Even if building a supercomputer only with commodity hardware, the design space is expanding significantly. 
This is because the number of available hardware choices is increasing dramatically because of the emerging architectures and device technologies, such as AI/graph/quantum/reconfigurable accelerators, 3D-stacked/persistent/processing-in memory or storage technologies, and smart/photonic interconnects, 
driven by the slowdown in Moore's law plus new types of workloads including HPDA and AI. 
Traditionally, the system configurations are determined in order to maximize performance of proxy applications while adhering to constraints like total budget, power supply, machine footprint, or weight. 
In the future,  system architects will need to take  \textit{carbon footprint budget} into account as another design constraint. 

They will have to assess the embodied carbon emissions for a variety of hardware devices and decide the system architecture so that the total embodied carbon footprint does not exceed the given limit.  
If this \textit{embodied carbon budget} is not fully used, the remaining part can be shifted to the \textit{operational carbon budget} in order to boost the system performance by raising the system power limit for a certain amount of time. 
\textit{Trading-off the embodied and operational carbon budgets under a total carbon footprint budget will be another optimization opportunity for  system designs. }
To enable this, we need reliable tools to accurately quantify the embodied carbon footprints without any bias. Further, once such tools exist, we should extend the existing supercomputing rankings to cover the carbon efficiency perspective (something like a \textit{Carbon500} list). 




\subsection{System Lifetime, Reuse, and Recycling}
\label{sec:lifetimerr}

Generally, the hardware refresh cycles for compute components and high-performance networking interconnects in HPC systems range between four and six years. For example, the lifetime values for the different HPC systems at LRZ~\cite{lrz} are summarized in Table~\ref{tab:hpcsystems}. In cloud data centers, decisions to retire existing servers are influenced by various factors such as maintenance costs and performance-per-watt metrics~\cite{mythsembodied}. In contrast, for public HPC centers, decommissioning of existing systems depends primarily on the preplanned duration of the project and the available funding. Therefore, while extending the lifetime of systems can significantly reduce embodied carbon emissions, it might not be feasible for public HPC data centers~\cite{serverlifetime, hyrax}. However, since storage and memory constitute a significant fraction of an HPC system's embodied carbon emissions (Fig.~\ref{fig:embodiedcarbon}), reusing existing components in the newer generation of HPC systems can significantly reduce embodied emissions. For instance, recent research targets reusing DDR4 memory chips from decommissioned servers in new DDR5 servers while maintaining performance~\cite{pond}. In addition, decommissioned servers from an HPC system can be relocated and reused in other educational institutions for teaching. For example, LRZ offers servers from decommissioned systems to other public institutions in Germany for free.

A common practice by public HPC centers is to include requirements in the Request for Proposal (RFP) that the vendor must take back and recycle the decommissioned servers. However, recycling yields relatively limited returns for reducing carbon emissions, while component reuse is significantly more effective~\cite{mythsembodied}. For example, resuing hard disk drives leads to $275$x more carbon emissions reductions than recycling. Recycling, however, is not insignificant and can provide several other benefits, such as recovering critical materials~\cite{frost2020use}. Finally, in terms of embodied carbon, server lifetime extensions are more effective than component reuse since not all server components can be effectively reutilized.







\begin{figure}[t]
\centering
\includegraphics[width=\columnwidth]{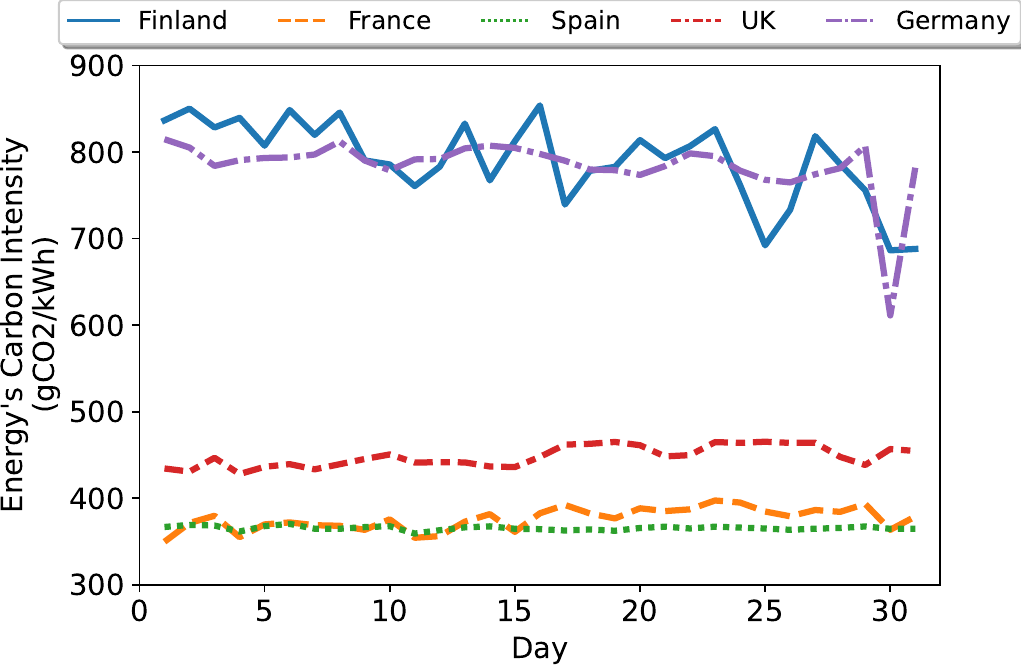}
\caption{Averaged daily marginal carbon intensities~\cite{marginal} for the different geographical regions across Europe in January 2023.}
\label{fig:opintensity}
\end{figure}

\section{Reducing Operational Carbon}
\label{sec:opcarbon}
Operational carbon refers to the carbon emissions produced during the day-to-day operation of large-scale HPC systems and data centers. It strongly depends on where the data center is located and the region's \textit{carbon intensity}. The carbon intensity of a geographical region describes the amount of carbon emitted per kWh of energy produced. The intensity can vary significantly across regions and even within the same region over time\footnote{Exceptions do exist, such as LRZ, which operates on a relatively constant carbon intensity due to agreements with the electricity provider (\S\ref{sec:embodiedcarbon}).}. For example, Fig.~\ref{fig:opintensity} shows the average daily carbon intensities for the different European counties in Jan. 2023, using a grid emissions data provider. In that month, Finland had $2.1$x higher carbon intensity compared to France. Moreover, the daily carbon intensity in Finland showed a standard deviation of $47.21$, highlighting temporal variability. As a result, depending on where an HPC center is situated, operational carbon can play a bigger role in its overall carbon impact. Opportunities for reducing operational carbon footprint can be divided into three categories: \circledw{1} designing carbon-aware system software (\S\ref{sec:rateshifting},\S\ref{sec:resourcescaling},\S\ref{sec:tempshifting}), \circledw{2} creating awareness and encouraging HPC users to reduce their carbon footprint (\S\ref{sec:quantifying emissions}), and \circledw{3} using energy-efficient hardware as described in \S\ref{sec:chip-level}.

\subsection{Carbon-aware Dynamic Power Budget Scaling}
\label{sec:rateshifting}
A holistic and extensible power management architecture was proposed by the HPC PowerStack community~\cite{powerstack}, and there exist several ongoing efforts to develop a production-level software stack to realize the concept in both academia and industry (e.g., EuroHPC's Regale project~\cite{regale}, READEX project~\cite{readex}).  
These power management approaches are typically based on a hierarchical and closed-loop control, as follows.
First, the site administrator inputs the total system power budget, and then the system management tool divides and distributes the given power budget accordingly to the currently running jobs. 
The given power budget is distributed across the allocated nodes for each job, and then the power budget at each node is split and assigned to the in-node hardware components (e.g., CPUs, GPUs, and DRAMs) by setting up their hardware knobs, typically power caps. 
This hierarchical power budgeting is supposed to operate based on hints, feedback by the users or monitoring tools because the optimal power budget setup depends highly on the dynamic characteristics of jobs.

These power management infrastructures are naturally extensible for reducing/limiting the operational carbon footprint as well. 
As the operational carbon footprint is the time integral of carbon intensity multiplied by power consumption, scaling up/down the total system power constraint in accordance with the carbon intensity changes (see Fig.~\ref{fig:opintensity}) is essential. 
This can be achieved by adding two properties to the PowerStack: a carbon intensity monitor and a simple mechanism to automatically determine the total system power budget based on it. 
As the total system power budget can significantly affect the job scheduling decision making, carbon intensity prediction can support the job scheduler, in particular when the system is setup for long running jobs. 
\subsection{Carbon-aware Dynamic Resource Scaling}
\label{sec:resourcescaling}
HPC resource management systems traditionally support only rigid or moldable jobs, i.e., they assign a fixed number of nodes statically when launching a job.
However, as various HPC algorithms and applications inherently are dynamic, supporting malleable jobs in batch schedulers, to which the node assignments are agnostic and dynamically changeable at runtime, has been studied~\cite{compres2016infrastructure,chadha2020extending,huber2022towards, chadha2020adaptive}. 
Malleability is a desired feature also for power-constrained systems~\cite{arima2022convergence}, as limiting the number of available nodes is an effective approach to keep the system under the given total power budget, which in turn can considerably change depending on the carbon intensity as discussed in \S\ref{sec:rateshifting}. To this end, the system manager and job manager in the PowerStack combined with a malleability supporting software stack should collaboratively and dynamically orchestrate (1) job power budget, (2) node allocation, and (3) power budget distributions across the allocated nodes simultaneously during runtime. 



\subsection{Carbon-aware Scheduling and Checkpointing}

\label{sec:tempshifting}
The fluctuating carbon intensity of the electricity grid creates \textit{green periods}, where the carbon intensity is significantly lower than the average carbon intensity for that location, as shown in Fig.~\ref{fig:opintensity}. To enhance the operating carbon efficiency of HPC systems, intelligent carbon-aware scheduling plugins for common resource and job management software (RJMS), such as \texttt{Flux} or \texttt{SLURM}, must be developed. Combined with forecasting techniques that leverage historical carbon intensity data, these plugins can intelligently backfill submitted jobs with suitable execution times during green periods. In addition, for long-running HPC jobs, carbon-aware checkpoint and restore strategies should be developed. These strategies can suspend the execution of the job during high carbon periods and resume execution when the intensity is low.


\subsection{Making HPC Users Greener}

\label{sec:quantifying emissions}
A common strategy employed by most HPC centers for efficient system management involves configuring multiple queues within the underlying RJMS software. These distinct queues are characterized by varying job scheduling priorities, constraints on the number of permissible nodes per job, and maximum job run times. This flexibility enables users to select the most appropriate queue according to the specific demands of their applications. However, our analysis of user job data from the SuperMUC-NG system reveals that many users allocate more nodes to their jobs than they require. 
This practice often leads to suboptimal utilization of system resources and contributes to higher carbon emissions~\cite{li2023sustainable}. To promote greater awareness among HPC users about the carbon impact of their jobs, it becomes important to provide them with carbon-related insights. To this end, it is necessary to extend operational data analytics tools, such as \texttt{DCDB}~\cite{dcdb}, to be able to quantify and aggregate carbon emissions data derived from submitted HPC jobs; only then a comprehensive HPC job carbon profile can be established and integrated into job reports, ensuring accessibility to HPC users. Moreover, the carbon footprint data can also be presented using \textit{analogies} that resonate with typical HPC system users. For example, by equating the emitted carbon to the carbon produced by driving a car between two regions within a country~\cite{suskeys}.

HPC centers commonly allocate compute budget to projects using units like \textit{core-hours}, enabling project members to execute HPC jobs. To encourage users to submit jobs during periods of green energy (Fig.~\ref{fig:opintensity}), HPC centers can offer incentives by only charging a fraction of the actual core hours used by the job during that time. This approach can be synergistically integrated with \S\ref{sec:tempshifting} to enable automatic incentivized HPC job budget accounting. In addition, users can proactively reduce the carbon footprint of their applications by utilizing application libraries such as Cesarini~et~al.~\cite{cesarini2020countdown} or by making their applications malleable (\S\ref{sec:resourcescaling}).

\section{Conclusion}
\label{sec:conclusionfuture}

In this position paper, we motivated the opportunities based on the quantitative data obtained in our region and highlighted our view on challenges for reducing both embodied and operational carbon footprints, including a wide range of aspects such as hardware design, system architecture, system life cycle, system software stack, and user encouragement.  

\begin{acks}
We thank Dr. Michael Ott for providing us his valuable insights and Philipp Friese for joining our early stage discussions. 
This work has received funding under the European Commission’s EuroHPC and Horizon 2020 programmes under grant agreements no. 956560 (REGALE), no. 955606 (DEEP-SEA), and no. 955701 (TIME-X).  
\end{acks}


{
\renewcommand{\baselinestretch}{0.1}
\bibliographystyle{ACM-Reference-Format}
\bibliography{serverless}
}

\end{document}